\documentclass[twocolumn,superscriptaddress,showpacs,aps,amsmath,amssymb,prb]{revtex4}

\newif\ifNotArXiv

\NotArXivfalse    	

\usepackage{graphicx}
\usepackage{subfigure}
\usepackage{color}
\ifNotArXiv
	\usepackage{todonotes}
	\usepackage{hyperref}
\fi

\newcommand{\ket}[1]{\left\vert #1 \right\rangle}
\newcommand{\bra}[1]{\left\langle #1 \right\vert}
\newcommand{\abs}[1]{\left\vert #1 \right\vert}
\newcommand{\erw}[1]{\left\langle #1 \right\rangle}
\newcommand{\komm}[2]{\left[ #1 , #2 \right]}

\renewcommand{\d}{\mathrm{d}}
\newcommand{\comma}{~,}
\newcommand{\fullstop}{~.}

\begin{document}

\title{Creating photon-number squeezed strong microwave fields by a Cooper-pair injection laser}

\author{Martin Koppenh\"ofer}
\affiliation{Department of Physics, University of Basel, CH-4056 Basel, Switzerland}
\affiliation{Institut f\"ur Theoretische Festk\"orperphysik, Karlsruhe Institute of Technology, D-76131 Karlsruhe, Germany}

\author{Juha Lepp\"akangas}
\affiliation{Institut f\"ur Theoretische Festk\"orperphysik, Karlsruhe Institute of Technology, D-76131 Karlsruhe, Germany}

\author{Michael Marthaler}
\affiliation{Institut f\"ur Theoretische Festk\"orperphysik, Karlsruhe Institute of Technology, D-76131 Karlsruhe, Germany}

\date{\today}


\begin{abstract}
The use of artificial atoms as an active lasing medium opens a way to construct novel sources of nonclassical radiation. 
An example is the creation of photon-number squeezed light. 
Here we present a design of a laser consisting of multiple Cooper-pair transistors coupled to a microwave resonator. 
Over a broad range of experimentally realizable parameters, this laser creates photon-number squeezed
microwave radiation, characterized by a Fano factor $F \ll 1$, at a very high resonator photon number. 
We investigate the impact of gate-charge disorder in a Cooper-pair transistor and show that the system can create squeezed strong microwave fields even in the presence of maximum disorder.
\end{abstract}

\maketitle

\section{Introduction}
Squeezed light is widely used in spectroscopic and interferometric experiments to enhance the measurement sensitivity
and to overcome the quantum shot-noise limit~\cite{Yuen1986,SqueezedLightGravitational,SpectroscopySqueezedLight,AgarwalScullyRamseySpectroscopy,MirrorMotionSqueezedLight,BookYamamoto,QubitReadoutSqueezedLight}. 
In low-temperature setups, such as in superconducting microwave circuits,
it could be a benefit to have a miniaturized source of nonclassical radiation,
which can be used for, e.g.~qubit control and measurement.

When driven by a coherent microwave drive, a transmission-line resonator terminated by a Josephson junction can be used to create
quadrature-squeezed microwave radiation~\cite{JBAReview,DeppeSqueezed,BeltramSqueezed,DrivenNonlinearOscillator,BJohansson2013}.
Without an external coherent drive, but with a simple DC-voltage bias, a
Josephson junction can be a source of microwaves with reduced photon-number fluctuations, so-called photon-number squeezed light~\cite{Leppakangas_Entangled,Paris2015,Ulm2015,Leppakangas2016,Clerk_Fock}.
Such an on-chip source of microwaves can also work as a traditional laser~\cite{LasingCassidy}.
It has also been shown that photon-number squeezing can be realized in lasing setups that use artificial atoms as an active medium if they provide a coupling to the radiation field that vanishes at certain photon numbers \cite{KM_Squeezing}.
Such couplings arise because of strong longitudinal $\sigma_z$ type coupling of qubits and semiconductor or gate defined quantum dots to the radiation field of a resonator. 
Single- or few-atom lasing in such setups has already been studied both theoretically and experimentally~\cite{SingleArtificialAtomLasing,SingleCooperPairJosephsonLaser,DoubleQuantumDotMaser,Lasing_Jin}, and similar couplings are found in setups that couple qubits to a resonator by a voltage-biased Josephson junction~\cite{MLC_Squeezed,KubalaSqueezedBiasedJunction,Clerk_Fock}.
However, in order to obtain a higher output power the number of artificial atoms has to be increased. 
The coupling of a large number of artificial atoms to a cavity has already been demonstrated experimentally~\cite{Metamaterial_Pascal,ManyFluxQubits}.

Lasing setups built out of superconducting qubits or quantum dots inevitably suffer from fluctuations of the lasing parameters due to an imperfect fabrication process and noise induced by the environment. 
However, it has been shown that conventional lasers producing coherent light are quite robust against such fluctuations~\cite{KMS_Disorder}.

In this article, we study a microwave laser that can create photon-number  squeezed light at very high photon numbers.
This is achieved by attaching multiple artificial atoms to a resonator through a coupling that vanishes at certain photon numbers. 
In contrast to quadrature squeezed light, photon-number squeezed light has reduced fluctuations in the radial direction in phase space~\cite{TeichSaleh}.
The proposed realization consists of Cooper-pair transistors\cite{SSET1,SSET2,SSET3,SSET4} (CPTs) connected to a microwave resonator.
This generalizes the coupling scheme of a single-atom laser introduced in Ref.~\onlinecite{MLC_Squeezed} to a multi-atom lasing setup.
We find that generation of photon-number squeezed states is possible up to very high photon numbers, $\erw{n} \gtrsim 10000$. 
We also find that the squeezing is robust against fluctuations of the energy-level splittings and the coupling strength that originate in gate-charge disorder of the CPTs.
In particular, we demonstrate that in a specific experimentally feasible parameter regime, squeezing at a high photon number can be achieved even in the presence of maximum gate-charge disorder.
It follows that this device is robust against strong thermal noise in the DC-voltage bias~\cite{Leppakangas2014} and nonequilibrium quasiparticles in the CPTs~\cite{QuasiJuha,QuasiHeimes,DeGraaf2013,DeGraaf2}.

The paper is organized as follows.
In Sec.~\ref{sec:System}, we introduce the Hamiltonian that contains a nonmonotonous coupling to the radiation field,
which is exploited to produce photon-number squeezed radiation.
We show that it can be realized by using CPTs coupled galvanically to a microwave resonator. 
In Sec.~\ref{sec:Lasing}, we introduce additional dissipative processes that are  needed to achieve lasing. 
We briefly review the analytical method that can be used to calculate the laser photon statistics for arbitrary photon numbers. 
In Sec.~\ref{sec:Squeezing}, we discuss our results on photon-number squeezing at very high photon numbers, with strong noise in the DC bias.
In Sec.~\ref{sec:Noise}, we analyze the influence of charge disorder, which can in certain limits be understood as very strong low-frequency fluctuations (such as $1/f$-noise) in the transistor gate charge.
We find a particular  regime, where squeezing is created even in the presence of maximal charge noise.

\section{The system}
\label{sec:System}
In this section, we introduce the Hamiltonian governing the coherent dynamics of the system.
We consider a circuit  consisting of multiple Cooper-pair transistors coupled to a microwave resonator.
In Sec.~\ref{sec:Lasing}, we map this Hamiltonian onto an effective lasing Hamiltonian and we introduce additional dissipative processes, which are needed to realize the pumping process necessary for microwave lasing.

\subsection{Circuit Hamiltonian}
\begin{figure}[t!]
	\centering
	\subfigure[~Circuit diagram]{
		\includegraphics[width=.48\textwidth]{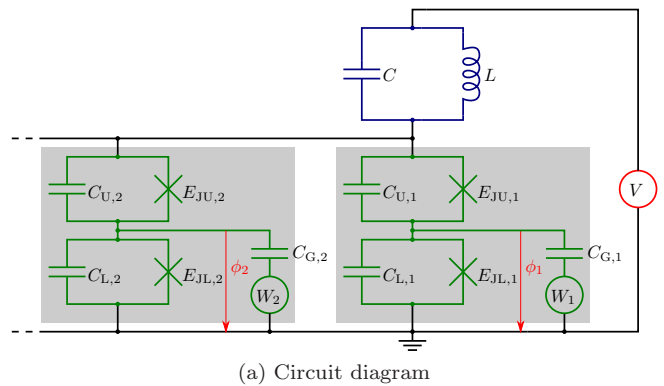}
		\label{fig:Setup:Diagram}
	}
	\subfigure[~Top-view sketch of the setup]{
		\includegraphics[width=.48\textwidth]{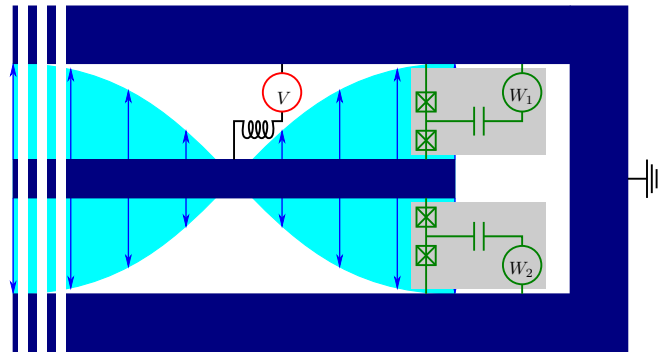}
		\label{fig:Setup:Sketch}
	}
	\caption{
		(a) Circuit diagram and (b) sketch of the lasing setup consisting of multiple superconducting Cooper-pair transistors (CPTs, gray boxes) that are coupled to a LC resonator. 
		A DC bias voltage $V$ is applied across the resonator and all CPTs.
		The voltage source is connected to a phase node of the resonator. Josephson junctions (crossed boxes) are modeled as Cooper-pair tunnel contacts (crosses) of coupling energies $E_{\mathrm{JU/JL},j}$ and parallel junction capacitances $C_{\mathrm{U/L},j}$. Each Cooper-pair transistor has an individual gate voltage $W_j$ through a gate capacitor $C_{\mathrm{G},j}$.
	}
	\label{fig:Setup:all}
\end{figure}

We consider a lasing setup consisting of $M$ Cooper-pair transistors coupled to a LC resonator with a DC-voltage bias $V$. 
The circuit diagram defining the system parameters is shown in Fig.~\ref{fig:Setup:Diagram}.
The sketch in Fig.~\ref{fig:Setup:Sketch} gives an example of an experimental realization of this coupling scheme. Similar schemes to realize a transport voltage across the resonator have already been proposed \cite{SingleCooperPairJosephsonLaser,SouquetNatCommun6562}.
Independent quantum-mechanical degrees of freedom of the setup are the superconducting phase differences $\phi_j$
across the lower Josephson junctions of the Cooper-pair transistors, $j \in \{1,\dots, M\}$, and the number $n$ of photons in the resonator.

The total Hamiltonian describing the coherent dynamics has the form
\begin{align}
	H = \hbar \omega_0 a^\dagger a + \sum_{j=1}^M H_{\mathrm{CPB},j} + \sum_{j=1}^M H_{\mathrm{int},j} \fullstop
	\label{eqn:Hcircuit}
\end{align}
The microwave resonator is modeled by the standard Hamiltonian term $\hbar \omega_0 a^\dagger a$, with $a^{(\dagger)}$
being the resonator photon annihilation (creation) operator and $\omega_0$ being its resonance frequency.
The dynamics of the phase $\phi_j$ on each superconducting island, situated between the upper and the lower Josephson junctions and the gate capacitors $C_{\mathrm{G},j}$, is described by a Cooper-pair box (CPB) Hamiltonian,
\begin{align}
	H_{\mathrm{CPB},j} = 4 E_{\mathrm{C},j} \left( N_j - N_{\mathrm{G},j} \right)^2 - E_{\mathrm{JL},j} \cos \left( \phi_j \right) \comma
\end{align}
which accounts for Cooper-pair tunneling across the lower Josephson junctions.
Note that this also corresponds to the Hamiltonian of a Transmon qubit.
The tunneling coupling across the upper junction is included in the interaction Hamiltonian, given below.
In this paper we operate in the charge regime and, therefore, the upper and the lower Josephson energies, $E_{\mathrm{JL},j}$ and $E_{\mathrm{JU},j}$, are assumed to be smaller than the charging energy $E_{\mathrm{C},j}$.
The operator $N_j$ counts the number of Cooper pairs on the superconducting island $j$ and it is the conjugate quantity to the phase difference $\phi_j$. 
It holds $[N_j,e^{\pm i\phi_j}]=\pm e^{\pm i \phi_j}$. 
The explicit expressions for the charging energy, the control gate charge $N_{\mathrm{G},j}$, and other parameters of the total Hamiltonian are given
in Sec.~\ref{sec:HamiltonianParameters}.

The interaction term of the total Hamiltonian consists of three different contributions,
\begin{align}
	H_{\mathrm{int},j} = H_{\mathrm{inj},j} + H_{\mathrm{cc},j} + H_{\mathrm{ic},j} \fullstop
	\label{eqn:InteractionHamiltonian}
\end{align}
The most important term for lasing is the ``injection'' term, describing Cooper-pair tunneling across the upper Josephson junctions.
It introduces a nonmonotonous coupling of the CPT to the resonator, 
\begin{align}
	H_{\mathrm{inj},j} = - E_{\mathrm{JU},j} \cos \left( \frac{2 e V t}{\hbar} + \mathcal{G} \left( a + a^\dagger \right) + \phi_j \right) \comma
\end{align}
which has roots at certain photon numbers.
A similar coupling (in the single-junction case) has been studied also in Refs.~\onlinecite{MLC_Squeezed,KubalaSqueezedBiasedJunction,Clerk_Fock}, and~\onlinecite{Ulm2013}.
The dimensionless coupling parameter $\mathcal{G}$ has the form
\begin{align}
	\mathcal{G} &= \sqrt{\frac{2 e^2}{\hbar} L \omega_0} \comma
\end{align}
which can equivalently be written as
\begin{align}
\mathcal{G}=\sqrt{\frac{\pi Z_{LC}}{R_{\rm Q}}} \comma
\end{align}
where
we have introduced the characteristic impedance of the resonator $Z_{LC}=\sqrt{L/\tilde C}$ and the quantum resistance $R_{\rm Q}=h/4e^2$.
The effective resonator capacitance $\tilde{C}$ is given in Sec.~\ref{sec:HamiltonianParameters}.

\subsection{Additional coupling terms}\label{sec:AdditionalCoupling}
Besides the injection interaction, additional coupling terms appear in the interaction Hamiltonian~\eqref{eqn:InteractionHamiltonian}.
First, there is a charge coupling (cc) of the CPTs to the resonator which is, however, off-resonant and will be neglected in the following using a rotating-wave approximation.
Second, there is an (ic) coupling between different CPTs. 
A mean field theory analysis (cf.\ Appendix~\ref{app:Justifications}) shows that for situations considered in this article
the contribution of the inter-CPT coupling is small and can be neglected.
For completeness, we also give the exact forms of the cc and ic interaction, 
\begin{align}
	H_{\mathrm{cc},j} &= - 2 i \frac{E_{\mathrm{cc},j}}{\mathcal{G}} N_j \left( a^\dagger - a \right) \comma 
	\label{eqn:AdditionalCouplings:CC}\\
	H_{\mathrm{ic},j} &= \sum_{\stackrel{l \neq j}{l=1}}^M 2 E_{\mathrm{ic},j,l} N_j N_l \fullstop
	\label{eqn:AdditionalCouplings:IC}
\end{align}
The coupling energies of these interaction terms are
\begin{align}
	E_{\mathrm{cc},j} &= \frac{e^2}{\tilde{C}} \frac{C_{\mathrm{U},j}}{C_{\Sigma,j}} \comma \\
	E_{\mathrm{ic},j,l} &= \frac{e^2}{\tilde{C}} \frac{C_{\mathrm{U},j}}{C_{\Sigma,j}} \frac{C_{\mathrm{U},l}}{C_{\Sigma,l}} \fullstop
\end{align}

\subsection{Hamiltonian parameters}\label{sec:HamiltonianParameters}
The numerical values of the parameters in the Hamiltonian~(\ref{eqn:Hcircuit}) are defined by the circuit variables introduced in Fig.~\ref{fig:Setup:Diagram}.
An important parameter is the resonator frequency $\omega_0$,
which has the form
\begin{align}
\omega_0 = \frac{1}{\sqrt{L \tilde{C}}} \fullstop
\end{align}
The effective resonator capacitance $\tilde{C}$ includes the bare resonator capacitance $C$ and additional normalization terms caused by the other capacitors,
\begin{align}
	\tilde{C} = C + \sum_{j=1}^M C_{\mathrm{U},j} - \sum_{j=1}^M \frac{C_{\mathrm{U},j}^2}{C_{\Sigma,j}} \fullstop
\end{align}
We introduced the usual abbreviation of the island capacitance $C_{\Sigma,j} = C_{\mathrm{U},j} + C_{\mathrm{L},j} + C_{\mathrm{G},j}$.
The charging energy of the superconducting island $j$ has the general form
\begin{align}
	E_{\mathrm{C},j} = \frac{e^2}{2} \frac{1}{\tilde{C}} \left( \frac{\tilde{C}}{C_{\Sigma,j}} + \frac{C_{\mathrm{U},j}^2}{C_{\Sigma,j}^2} \right) \comma
\end{align}
and the corresponding control gate charge $N_{\mathrm{G},j}$ is 
\begin{align}
	N_{\mathrm{G},j} = \frac{1}{2 e} \Bigg[ \begin{aligned}[t] 
		- &\frac{C C_{\mathrm{U},j}}{\tilde{C} + \frac{C_{\mathrm{U},j}^2}{C_{\Sigma,j}}} V + C_{\mathrm{G},j} W_j \\
		+ &\frac{C_{\mathrm{U},j}}{\tilde{C} + \frac{C_{\mathrm{U},j}^2}{C_{\Sigma,j}}} \sum_{\stackrel{l \neq j}{l=1}}^M \frac{C_{\mathrm{U},l} C_{\mathrm{G},l}}{C_{\Sigma,l}} W_l \Bigg] \fullstop \end{aligned} 
\end{align}

\section{Lasing Hamiltonian and dissipative processes}\label{sec:Lasing}
In this section, we map the circuit Hamiltonian~\eqref{eqn:Hcircuit} onto an effective Hamiltonian describing the coherent lasing interaction between the two states of the lasing transition. 
We restrict the Cooper-pair box Hilbert space to a two-level system, apply a rotating-wave approximation, and couple an additional dissipative environment to the coherent system, which is needed to induce population inversion and microwave lasing.
We neglect the effect of the cc and ic coupling terms listed in Sec.~\ref{sec:AdditionalCoupling}, which are in our case not relevant for lasing. 
For the sake of completeness, the full form of the Hamiltonian in a rotating wave approximation, including the additional coupling terms caused by the cc and ic interactions, is given in the Appendix~\ref{sec:FullHamiltonian}.

\subsection{Lasing Hamiltonian}
Each Cooper-pair box  is operated close to its symmetry point $N_{\mathrm{G},j} = 1/2$,
where the lowest transistor charge states $\ket{N_j=0}$ and $\ket{N_j=1}$ are degenerate. 
We restrict the CPB Hilbert space to these two states and diagonalize the resulting Hamiltonian by introducing the two-level basis
\begin{align}
	\ket{\uparrow_j} &= \cos \left( \frac{\theta_j}{2} \right) \ket{1_j} + \sin \left( \frac{\theta_j}{2} \right) \ket{0_j} \comma \\
	\ket{\downarrow_j} &= \sin \left( \frac{\theta_j}{2} \right) \ket{1_j} - \cos \left( \frac{\theta_j}{2} \right) \ket{0_j} \comma
\end{align}
where the mixing angle $\theta_j$ is defined by
\begin{align}
	\tan (\theta_j) &= - \frac{E_{\mathrm{JL},j}}{4 E_{\mathrm{C},j} (1 - 2 N_{\mathrm{G},j})} \fullstop
	\label{eqn:MixingAngle}
\end{align}

Next, we switch to a rotating frame by an unitary transformation $U = \exp ( - 2 i e V a^\dagger a t / \hbar)$ and we perform a rotating wave approximation.
The resulting effective lasing Hamiltonian reads
\begin{align}
	H_\mathrm{eff} &= \hbar \omega_\mathrm{eff} a^\dagger a + \sum_{j=1}^M \frac{\epsilon_j}{2} \sigma_z^j \nonumber \\
	& + \sum_{j=1}^M \sum_{n=0}^\infty \left( \hbar A_{n+1,n}^j \sigma_+^j \ket{n+1} \bra{n} + \mathrm{h.c.} \right) \fullstop
	\label{eqn:Heff}
\end{align}
Here $\sigma_{z}^j$ is the Pauli matrix acting on the subspace of the $j$-th Cooper-pair box.
The matrices $\sigma_{+}^j$ and $\sigma_{-}^j$ are the corresponding atomic raising and lowering operators, $\sigma_\pm^j = (\sigma_x^j \pm i \sigma_y^j)/2$.
The level-splitting energy is 
\begin{align}
 	\epsilon_j = \sqrt{16 E_{\mathrm{C},j}^2 (1 - 2 N_{\mathrm{G},j})^2 + E_{\mathrm{JL},j}^2} \fullstop
 	\label{eqn:LevelSplittingEnergy}
\end{align}
We have chosen the effective resonator frequency
\begin{align}
	\omega_\mathrm{eff} = \omega_0 - 2 e V/\hbar 
	\label{eqn:EffectiveFrequency}
\end{align}
to be negative. 
Therefore the interaction term, given by the last term on the right-hand side of Eq.~(\ref{eqn:Heff}),
simultaneously excites the two-level system and puts one extra photon into the resonator.
The resonance condition has the form  $\epsilon_j = \hbar \abs{\omega_\mathrm{eff}} = 2 e V - \hbar \omega_0$.

The coupling matrix elements $A_{n+1,n}^j$ can be expressed in terms of the generalized Laguerre polynomials $L_n^m(x)$~\cite{Wunsche1991},
\begin{align}
	A_{n+1,n}^j 
		&= - \frac{E_{\mathrm{JU},j}}{2 \hbar} \sin^2 \left( \frac{\theta_j}{2} \right) \bra{n+1} e^{- i \mathcal{G} (a + a^\dagger )} \ket{n} \nonumber \\
		&= \frac{E_{\mathrm{JU},j}}{2 \hbar} \sin^2 \left( \frac{\theta_j}{2} \right) \frac{i \mathcal{G} e^{- \mathcal{G}^2/2}}{\sqrt{n+1}} L_n^1 (\mathcal{G}^2) \fullstop 
		\label{eqn:CouplingMatrixElements}
\end{align}
In the limit $\mathcal{G} \to 0$ the matrix elements reduce to the usual Jaynes-Cummings type coupling, $\vert A_{n+1,n}^j \vert \propto E_{\mathrm{JU},j} \mathcal{G} \sqrt{n+1}$.
However, for a nonzero value of $\mathcal{G}$ the Laguerre polynomials introduce oscillations of $A_{n+1,n}^j$ as a function of the photon number $n$.
The coupling matrix elements (almost) vanish at certain photon numbers, which we call ``roots of the coupling matrix elements'' in the following. 
The plot of the dimensionless coupling matrix elements $\tilde{A}_{n+1,n}^j = A_{n+1,n}^j \hbar/(E_{\mathrm{JU},j} \mathcal{G})$ in the upper inset of Fig.~\ref{fig:Squeezing} shows that the position of these roots of the coupling is controlled by the parameter $\mathcal{G}$:
For a smaller value of $\mathcal{G}$ the roots are situated at larger photon numbers.

In the effective lasing Hamiltonian~\eqref{eqn:Heff} we neglected multi-photon transitions. 
In general the corresponding matrix elements $A_{n+m,n}^j$ for $m \neq 1$ are nonzero and these multi-photon transitions may drive the laser across the squeezing point. 
However, they are off-resonant and their transition rates are suppressed compared to the dominating single-photon transition rates by a factor of $\mathcal{G} \Gamma_\varphi^2/\omega_\mathrm{eff}^2$, which is much smaller than unity for typical lasing parameters.

\subsection{Lasing processes}
The physical processes behind lasing in our system are the following.
When a Cooper pair tunnels across the upper Josephson junction into the resonator (is ``injected'' into the resonator), the energy $2 e V$ is gained, which is used to excite the two-level system and to emit a photon into the resonator. 
Population inversion is established by a relaxation of the two-level system into its ground state, i.e., the pumping process of atom $j$ is given by
\begin{align*}
	\ket{\downarrow_j,n} \stackrel{\text{C.p. tunnels}}{\longrightarrow} \ket{\uparrow_j,n+1} \stackrel{\text{TLS relaxes}}{\longrightarrow} \ket{\downarrow_j, n+1} \dots 
\end{align*}
The relaxation process emerges from the coupling of the Cooper-pair transistors to a dissipative environment, whose physical origin and modeling is discussed in the following section.

\subsection{Quantum master equation}
Lasing is modeled by an effective quantum master equation for the density matrix $\rho$ of the resonator and the artificial two-level atoms~\cite{Scully},
\begin{align}
	\frac{\d}{\d t} \rho = - \frac{i}{\hbar} \komm{H_\mathrm{eff}}{\rho} + \mathcal{L}_\mathrm{res} \rho + \sum_{j=1}^M \mathcal{L}_{\mathrm{at},j} \rho \comma
	\label{eqn:QuantumMasterEquation}
\end{align}
where $H_\mathrm{eff}$ is given by Eq.~\eqref{eqn:Heff}. 
The resonator decay is accounted for by a Lindblad superoperator
\begin{align}
	\mathcal{L}_\mathrm{res} \rho &= \frac{\kappa}{2} \left( 2 a \rho a^\dagger - a^\dagger a \rho - \rho a^\dagger a \right) \fullstop
\end{align}
where $\kappa$ is the resonator decay rate. 
For instance, it accounts for a coupling of the resonator to a transmission line.
The properties of the created microwave radiation in the transmission line can be measured in an experimental setup
and are directly related to the state of the resonator~\cite{WallsMilburn}.

Furthermore, we introduced the decay rate $\Gamma_{\downarrow,j}$ and the pure dephasing rate $\Gamma_{\varphi,j}^*$ of the artificial atom, accounted for by a Lindblad superoperator
\begin{align}
	\mathcal{L}_{\mathrm{at},j} \rho &= \frac{\Gamma_{\downarrow,j}}{2} \left( 2 \sigma_-^j \rho \sigma_+^j - \rho \sigma_+^j \sigma_-^j - \sigma_+^j \sigma_-^j \rho \right) \nonumber \\
	&+ \frac{\Gamma_{\varphi,j}^*}{2} \left( \sigma_z^j \rho \sigma_z^j - \rho \right) \fullstop 
\end{align}
The atomic decay rate $\Gamma_{\downarrow,j}$ plays the role of a pumping rate of the laser.
Atomic decay can be induced through high-frequency (quantum) fluctuations of the control voltages~\cite{MLC_Squeezed}.
If realized in this way, its magnitude can be engineered by the shape of the impedance of the biasing circuit:
Dissipation can be increased at desired frequencies, e.g., by connecting the Cooper-pair transistors to a low-quality-factor resonator~\cite{Clerk_Fock}.

The phenomenological pure dephasing rate $\Gamma_{\varphi,j}^*$ models low-frequency charge noise
in circuit-QED setups~\cite{chargeNoise1,chargeNoise2} and it also accounts for low-frequency noise in the bias voltage
due to thermal fluctuations~\cite{SSET4}.
For a $50\,\Omega$ low-frequency impedance at $T=20\,\mathrm{mK}$, thermal fluctuations contribute with
$\Gamma_{\varphi,j}^*=2\pi\times 20\,\mathrm{MHz}$~\cite{Leppakangas2014}.

A Lindblad-type master equation approach like Eq.~\eqref{eqn:QuantumMasterEquation} cannot account for very strong
low-frequency fluctuations, such as gate-charge $1/f$ noise. 
In this case other types of master equations can be
used~\cite{Marthaler2016}. 
We analyze the influence of this type of noise as quasistatic ``charge disorder'', as shown in Sec.~\ref{sec:Noise}.

\subsection{Solution method}\label{sec:SolutionMethod}
We obtain the laser photon statistics $p(n)$ out of the quantum master equation~\eqref{eqn:QuantumMasterEquation} using the approach described in Ref.~\onlinecite{KM_Squeezing}. 
It amounts to derive a recursion relation of the laser photon statistics, which can be solved numerically:
\begin{align}
	p(n) = \frac{p(n-1)}{\kappa n} \sum_{j=1}^M \frac{ 2 \Gamma_{\downarrow,j}^2 \Gamma_{\varphi,j} \vert A_{n,n-1}^j \vert^2}{M_n^j}  .
	\label{eqn:Lasing:PhotonStatisticsRecursionRelation}
\end{align}
Here we introduced the total dephasing rate
\begin{equation}
\Gamma_{\varphi,j} = \Gamma_{\downarrow,j}/2 + \Gamma_{\varphi,j}^*,
\end{equation}
the detuning
\begin{equation}
\Delta_j = \epsilon_j/\hbar - \abs{\omega_\mathrm{eff}},
\end{equation}
and the denominator
\begin{eqnarray}
M_n^j = \Gamma_{\downarrow,j}^2 \left[ \Delta_j^2 + \Gamma_{\varphi,j}^2 \right] + 4 \Gamma_{\downarrow,j} \Gamma_{\varphi,j} \vert A_{n,n-1}^j \vert^2.
\end{eqnarray}
This solution is based on an adiabatic decoupling of the resonator and atomic degrees of freedom, i.e., the time scale of changes of the resonator state is much smaller than the timescale of changes of the atomic state, $\kappa \ll \Gamma_{\downarrow,j}, \Gamma_{\varphi,j}^*$. 
For very small photon numbers, $n \lesssim 100$, the quantum master equation~\eqref{eqn:QuantumMasterEquation} can also be solved numerically without applying an adiabatic elimination. 
The effect of photon-number squeezing discussed below is observed also in this numerical approach.
The recursion relation~\eqref{eqn:Lasing:PhotonStatisticsRecursionRelation} and the normalization condition $\sum_{n=0}^\infty \rho(n) = 1$ define the photon statistics uniquely. 
The solution is constructed by starting at an arbitrary initial photon number $n_0$ and an arbitrary initial value of $p(n_0)$ and then normalizing. 
In order to improve the numeric stability we take the logarithm of Eq.~\eqref{eqn:Lasing:PhotonStatisticsRecursionRelation} and evaluate it as a sum instead of as a product of iterative terms.
Within this method we can rederive the results of Ref.~\onlinecite{MLC_Squeezed} ($M=1$)
and extend this analysis to very high photon numbers by considering the case $M \gg 1$.

\begin{figure}[tb]
	\centering
	\includegraphics[width=.48\textwidth]{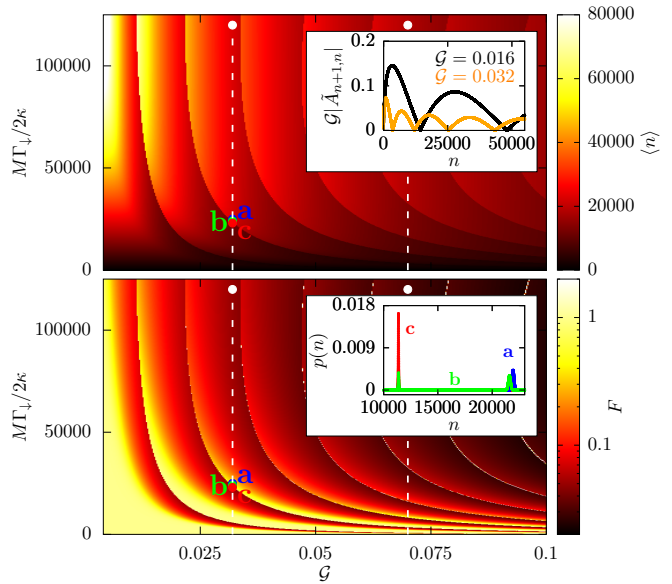}
	\caption{
		Photon-number expectation value $\erw{n}$ (upper main plot) and Fano factor $F$ (lower main plot) as a function of the dimensionless coupling parameter $\mathcal{G}$ and the photon number of a conventional laser, $M \Gamma_\downarrow/2 \kappa$, for a lasing setup consisting of $M=200$ identical CPTs. 
		The upper inset displays the modulus of the dimensionless coupling matrix elements $\tilde{A}_{n+1,n}$ for two different values of $\mathcal{G}$, which determines the position of the roots of $\tilde{A}_{n+1,n}$.
		The lower inset displays the photon statistics $p(n)$ of the laser for three different effective pumping rates $\Gamma_\downarrow = 0.0125, 0.012008, 0.0115\,\abs{\omega_\mathrm{eff}}$ and $\mathcal{G} = 0.032$, which are marked by labeled dots along the dashed white line in the main plots. 
		Plot parameters are $\kappa = 5 \times 10^{-5} \,\abs{\omega_\mathrm{eff}}$, $\Delta = 0$, $\theta = - \pi/2$, $E_{\mathrm{JU}} = 0.5\,\hbar\abs{\omega_\mathrm{eff}}$, and $\Gamma_\varphi^* = 0.013\,\abs{\omega_\mathrm{eff}}$. 
		The CPT relaxation rate has been varied in the range $5 \times 10^{-5}\,\abs{\omega_\mathrm{eff}} \leq \Gamma_\downarrow \leq 0.0625\,\abs{\omega_\mathrm{eff}}$. 
		The resonator frequency is $\omega_0 = 1.32 \,\abs{\omega_\mathrm{eff}}$.  
	}
	\label{fig:Squeezing}
\end{figure}

\section{Photon-number squeezing}\label{sec:Squeezing}
In this section, we present numerical results for the steady-state photon-number expectation value, $\erw{n}= \erw{a^\dagger a}$, and the Fano factor, 
\begin{align}
	F = \frac{\erw{n^2} - \erw{n}^2}{\erw{n}} .
\end{align}
We consider a setup consisting of $M$ CPTs with identical parameters. 
Lasing in the presence of varying parameters is discussed in Sec.~\ref{sec:Noise}. 
We study the dependence of $\erw{n}$ and $F$ on the dimensionless coupling parameter $\mathcal{G}$,
the Josephson energy $E_\mathrm{JU}$, the resonator decay rate $\kappa$, the number of atoms $M$, and the atomic relaxation rate $\Gamma_\downarrow$.

\subsection{Photon number squeezing vs. conventional lasing}
\label{sec:Lasing:PhotonNumberSqueezing}
Fig.~\ref{fig:Squeezing} shows plots of the photon-number expectation value $\erw{n}$ and the Fano factor $F$ as a function of the dimensionless coupling parameter $\mathcal{G}$ and the photon number of a conventional laser at resonance,
\begin{align}
	n_\mathrm{c} = M \Gamma_\downarrow/2 \kappa \fullstop
	\label{eqn:Squeezing:PhotonNumberExpValueConventional}
\end{align}
The results are obtained by a numerical solution of the recursion relation~\eqref{eqn:Lasing:PhotonStatisticsRecursionRelation}. 
To discuss the lasing behavior it is convenient to rewrite the recursion relation for $M$ identical resonant atoms as follows,
\begin{align}
	p(n) = \frac{p(n-1)}{n} \frac{n_\mathrm{c}}{1 + \frac{n_\mathrm{c}}{C \abs{\tilde{A}_{n,n-1}}^2}} \comma
	\label{eqn:RecursionRelationRephrased}
\end{align}
where we introduced the cooperativity parameter $C$~\cite{GardinerQuantumNoise}, 
\begin{align}
	C &= 2 \frac{M}{\kappa} \frac{E_\mathrm{JU}^2  \mathcal{G}^2}{\hbar^2 \Gamma_\varphi} \comma
	\label{eqn:Squeezing:Cooperativity}
\end{align}
and $\tilde{A}_{n,n-1} = A_{n,n-1}\hbar/(E_\mathrm{JU} \mathcal{G})$ is the dimensionless coupling matrix element. 
If the product $C \vert \tilde{A}_{n,n-1} \vert^2$ is large compared to $n_\mathrm{c}$, i.e., if the condition $\abs{A_{n,n-1}}^2 \gg \Gamma_\downarrow^2$ holds, the system behaves like a conventional laser and is governed by the recursion relation
\begin{align}
	p(n) = \frac{n_\mathrm{c}}{n} p(n-1) \fullstop
	\label{eqn:Squeezing:RecursionRelationConventional}
\end{align}
Then, the photon-number expectation value $\erw{n} \approx n_\mathrm{c}$ scales linearly with the effective pumping rate $M \Gamma_\downarrow$.
We observe a Poissonian photon statistics, characterized by a Fano factor close to unity, $F \approx 1$ (yellow (bright) areas in the lower main plot of Fig.~\ref{fig:Squeezing}).
In particular, this regime is realized for very small values of $\mathcal{G}$. 

The interesting squeezing effect occurs in the opposite limit, $\abs{A_{n,n-1}}^2 \ll \Gamma_\downarrow^2$, which is realized if the photon-number expectation value gets close to a root of the coupling matrix element $A_{n,n-1}$. 
In this regime, the recursion relation takes the form
\begin{align}
	p(n) = C \frac{\vert\tilde{A}_{n,n-1} \vert^2}{n} p(n-1) \fullstop
	\label{eqn:Squeezing:RecursionRelationTrapped}
\end{align}

Because the coupling of the atoms to the resonator breaks down close to a root of the coupling matrix elements, the photon number takes an almost constant value even if the effective pumping strength is increased. 
The region below point ``c'' in Fig.~\ref{fig:Squeezing} is an example of such a regime where the photon-number is ``trapped'' by a root of the coupling matrix element.
Deep in this regime, fluctuations of the photon number are strongly suppressed and a photon-number squeezed (sub-Poissonian) photon statistics, characterized by a Fano factor $F \ll 1$, is obtained:
An intuitive explanation is that if the photon number decreases because of losses, the coupling of the atom to the resonator becomes nonzero again and the strongly-pumped atoms quickly compensate the loss of photons. 

For even stronger pumping the trapping by the root is overcome and conventional lasing is observed (e.g.\ point ``a") until the photon number gets close to the position of the next root. 
At the transition from a regime where the photon-number expectation value is trapped by a root of the coupling to a regime of conventional lasing, a double-peaked photon statistics is observed, corresponding to a very large Fano factor $F \gg 1$ (point ``b").
There, the system exhibits multi-stability~\cite{MLC_Squeezed}.
The lower inset of Fig.~\ref{fig:Squeezing} shows the change of the photon statistics from a state trapped at a root of the coupling matrix element (``c'') via a multistable state with a double-peaked photon statistics (``b'') to a conventional lasing state (``a'').

\subsection{Role of the lasing parameters}
In the idealized system of identical CPTs, the photon-number expectation value $\erw{n}$ corresponds either to the conventional photon number $n_\mathrm{c}$, defined by Eq.~\eqref{eqn:Squeezing:PhotonNumberExpValueConventional}, or it is trapped close to the photon number of the highest root of the coupling matrix element which is still smaller than $n_\mathrm{c}$. 
Given a fixed atomic relaxation rate $\Gamma_\downarrow$, the value of $n_\mathrm{c}$ can be modified by the number of atoms $M$ and the resonator decay rate $\kappa$, which are, however, both subjected to physical constraints.
For example, engineering an extremely small resonator decay rate $\kappa$  in a setup with a sufficiently large atomic
relaxation rate $\Gamma_\downarrow$, needed for the pumping, might be difficult since they both correspond to
two different frequency responses of the same nearby electromagnetic environment.
The number of atoms $M$ is restricted by the physical size of an artificial atom compared to the resonator size.

The parameter $\mathcal{G}$ determines the position of the roots of the coupling matrix elements (cf.\ upper inset in Fig.~\ref{fig:Squeezing}) and it influences the coupling strength of the atoms to the resonator, $E_\mathrm{JU} \mathcal{G}$, cf.\ Eq.~\eqref{eqn:CouplingMatrixElements}.
As discussed in Sec.~\ref{sec:Noise}, in a disordered setup a small value of $\mathcal{G}$ is preferred in order to have a high laser intensity that is robust against disorder.
However, arbitrary small values of $\mathcal{G}$ are not realistic in an experimental realization.
An experimentally feasible characteristic impedance of the resonator is
$Z_{LC}=10\,\Omega$, which corresponds to $\mathcal{G}= 0.07$. 
This is indeed sufficient to create strongly squeezed fields with high photon numbers, as shown in Sec.~\ref{sec:Noise}.
Lower values of $\mathcal{G}$ could be reached, for example, by placing Cooper-pair transistors
closer to a phase node of the resonator, see Fig.~\ref{fig:Setup:Sketch}.

For a conventional laser at resonance, the coupling energy $E_\mathrm{JU}$ and the coupling parameter $\mathcal{G}$ determine the lasing threshold, where the cooperativity parameter~\eqref{eqn:Squeezing:Cooperativity} takes the value of unity. 
Far above the lasing threshold, $C \gg 1$, the photon number of a conventional laser, given by Eq.~\eqref{eqn:Squeezing:PhotonNumberExpValueConventional}, is independent of $E_\mathrm{JU}$ and $\mathcal{G}$. 
The parameters of Fig.~\ref{fig:Squeezing} are chosen such that the cooperativity $C$ stays 
in this regime in the entire range of values of $\mathcal{G}$.
However, in the squeezed regime and for a fixed value of $\mathcal{G}$, the coupling energy $E_\mathrm{JU}$  influences the lasing behavior far above the lasing threshold, because it determines the width of the trapping regions, defined by the relation $E_\mathrm{JU}^2 \mathcal{G}^2 \vert \tilde{A}_{n,n-1} \vert^2/\hbar^2 \ll \Gamma_\downarrow^2$.  A larger coupling energy leads to a reduced trapping region. 

Equation~\eqref{eqn:RecursionRelationRephrased} indicates that Fig.~\ref{fig:Squeezing} also describes the lasing behavior of other setups if the lasing parameters are rescaled such that the cooperativity $C$ is kept constant. 

Note that, contrary to the model discussed in Ref.~\onlinecite{KM_Squeezing}, already the coupling matrix elements of the initial circuit Hamiltonian~\eqref{eqn:Hcircuit} have roots, hence there is no need for a polaron transformation.
Therefore, the number $M$ of atoms in the setup does not impose a lower bound on the achievable Fano factor.
Vice versa, in this Cooper-pair transistor lasing setup, the squeezing and the brightness  of the laser can be adjusted independently.

\subsection{Schemes for experimental detection}
The created photonic states can be probed directly experimentally when the resonator decay is implemented by a coupling
to a nearby transmission line, which guides the emitted radiation to microwave detectors.
In this case, the flux of the outgoing photons, $f(t)$, maps onto the photon number in the resonator,
\begin{equation}
f(t)=a_{\rm out}^\dagger(t)a_{\rm out}(t)=\kappa n(t) \, .
\end{equation}
The photon flux can be measured by intensity or linear detectors~\cite{DaSilva2010}.

A closely related, and recently actively studied quantity in the microwave regime, is the second-order coherence
\begin{align}
	g^{(2)}(t) = \frac{\erw{a^\dagger a ^\dagger(t) a(t) a}}{\erw{n}^2}\, .
\end{align}
We assume here that $\erw{n}=\erw{a^\dagger a}$ is a constant.
Inside the cavity, this is related to the discussed Fano factor as
\begin{align}
	g^{(2)}(0) = 1 + \frac{F - 1}{\erw{n}} \, .
\end{align}
One sees that if we have a squeezed state, $F<1$, we get $g^{(2)}(0)<1$, i.e., the field is antibunched
and thereby non-classical~\cite{WallsMilburn}.
For a strongly squeezed state, $F \to 0$, the second-order coherence approaches the value for Fock states,
$g^{(2)}_\mathrm{Fock}(0) = 1 - 1/\erw{n}$. 
Vice versa, if we have $F > 1$, such as in the multi-stable situation, the statistics of the 
photons is bunched, $g^{(2)}(0)>1$.

The corresponding relation for the propagating photons reads~\cite{WallsMilburn}
\begin{align}\label{eq:IntensityFluctuations}
	\erw{f(t)f(0)} -\erw{f}^2 = \erw{f}\delta(t) + \erw{f}^2\left[ g^{(2)}(t)-1 \right] \, .
\end{align}
Here, the second-order coherence $g^{(2)}(t)$ has the same value inside and outside the resonator.
This relation is valid when the detector bandwidth is infinite or much larger than $\kappa$.
The first term on the right-hand side of Eq.~(\ref{eq:IntensityFluctuations}) corresponds to the photon shot noise. The second term is zero for a coherent
field, since we have $g^{(2)}(t)=1$ in this case.
Therefore, the intensity fluctuations of coherent radiation are characterized by shot noise.
On the other hand, for a Fock state in the resonator, we have $g^{(2)}(0)=1-1/\erw{n}$.
It is reasonable to assume that this decays towards one in a characteristic time defined by the inverse bandwidh of the resonator: $g^{(2)}(t)-1=e^{-\kappa t}/\erw{n}$.
In this case, we have
\begin{align}
	\int_0^\infty d t \left[ \erw{f(\tau)f(0)} -\erw{f}^2\right] = 0  \, ,
\end{align}
since the relation $\erw{f}=\kappa\erw{n}$ holds.
We then obtain the result that the
radiation from a perfectly photon-number squeezed resonator produces a photon flux (or intensity) that has no zero-frequency noise.

\section{Effect of charge disorder}
\label{sec:Noise}
Lasing setups based on artificial-atom technology are inevitably subjected to fluctuations in the atomic parameters, either due to deviations in the fabrication process or due to a coupling to the environment.
This yields, for instance, quasistatic fluctuations of the gate charge $N_{\mathrm{G},j}$~\cite{chargeNoise1,chargeNoise2}.
It is important to notice our discrimination between quasistatic charge noise and artificial-atom dephasing.
The former can be a very strong low-frequency effect, which cannot be accounted for by a Lindblad-type approach.
The latter is a weaker and usually faster, but still low-frequency phenomenon and is in our case dominated by thermal fluctuations of the bias voltage. 
It is accounted for within the Lindblad-type approach discussed in Sec.~\ref{sec:Lasing}.

In this section, we analyze the effect of strong quasistatic charge noise.
In the limit of maximum disorder, it is also a rough model for (non-equilibrium) quasiparticle poisoning in
the CPTs, which can be triggered in the high-intensity limit by quasiparticle tunneling due to direct multi-photon absorption~\cite{DeGraaf2013}.

For conventional lasers it has been shown that the lasing state is robust against disorder in the atomic detuning, the coupling strength to the resonator, and the pumping strength \cite{KMS_Disorder}. 
However, photon-number squeezing is sensitive to disorder as it relies on the existence of a root of the coupling matrix elements at the same photon number for all atoms in the setup. 

Contrary to other proposals, e.g., Ref.~\onlinecite{KM_Squeezing}, all atoms in the CPT lasing setup have their roots of the coupling at exactly the same photon numbers because in Eq.~\eqref{eqn:CouplingMatrixElements} the parameter $\mathcal{G}$ is identical for all atoms and it only depends on all fixed capacitances of the setup. 
Therefore, a reduction of squeezing due to different positions of the roots of the atomic coupling is not expected.

\subsection{Model of charge disorder}
The most significant remaining source of disorder is charge noise, i.e., fluctuations of the gate charge $N_{\mathrm{G},j}$ of the Cooper-pair transistors. 
According to Eqs.~\eqref{eqn:MixingAngle} and~\eqref{eqn:LevelSplittingEnergy} it causes fluctuations of the mixing angle $\theta_j$ and the level-splitting energy $\epsilon_j$. 
Therefore, the coupling strength of the lasing interaction, cf.\ Eq.~(\ref{eqn:CouplingMatrixElements}), and the detuning $\Delta_j = \epsilon_j/\hbar - \abs{\omega_\mathrm{eff}}$ fluctuate.

We model charge noise in the setup by a box distribution of width $b$, which is centered at the mean gate charge $\overline{N}_\mathrm{G} = 1/2$,
\begin{align}
	p_\mathrm{B}(N_\mathrm{G};b) &= \frac{1}{b} \left[ \Theta \left( N_\mathrm{G} - \frac{1}{2} + \frac{b}{2} \right) - \Theta \left( N_\mathrm{G} - \frac{1}{2} - \frac{b}{2} \right) \right] \comma 
\end{align}
where $\Theta$ is the Heaviside step function.
Such a box distribution has been used to model charge noise in previous studies~\cite{Marthaler_box_1,Marthaler_box_2}.
According to the box distribution $p_\mathrm{B}(N_\mathrm{G};b)$, a random gate charge $N_{\mathrm{G},j}$ is chosen for each CPT in the $M$ atom lasing setup. 
Using this set of random gate charges, the individual atomic mixing angles and level-splitting energies are obtained using Eqs.~\eqref{eqn:MixingAngle} and~\eqref{eqn:LevelSplittingEnergy} and the photon-number expectation value $\erw{n}$ and the Fano factor $F$ are calculated numerically.  
To obtain the mean photon-number expectation value $\overline{\erw{n}}$ and the mean Fano factor $\overline{F}$ we average the results of $N_\mathrm{runs}= 500$ randomly generated realizations of such a disordered multi-atom lasing system.

\subsection{Results}
\begin{figure}[tb]
	\centering
	\subfigure[~$\mathcal{G}=0.032$]{
		\includegraphics[width=.48\textwidth]{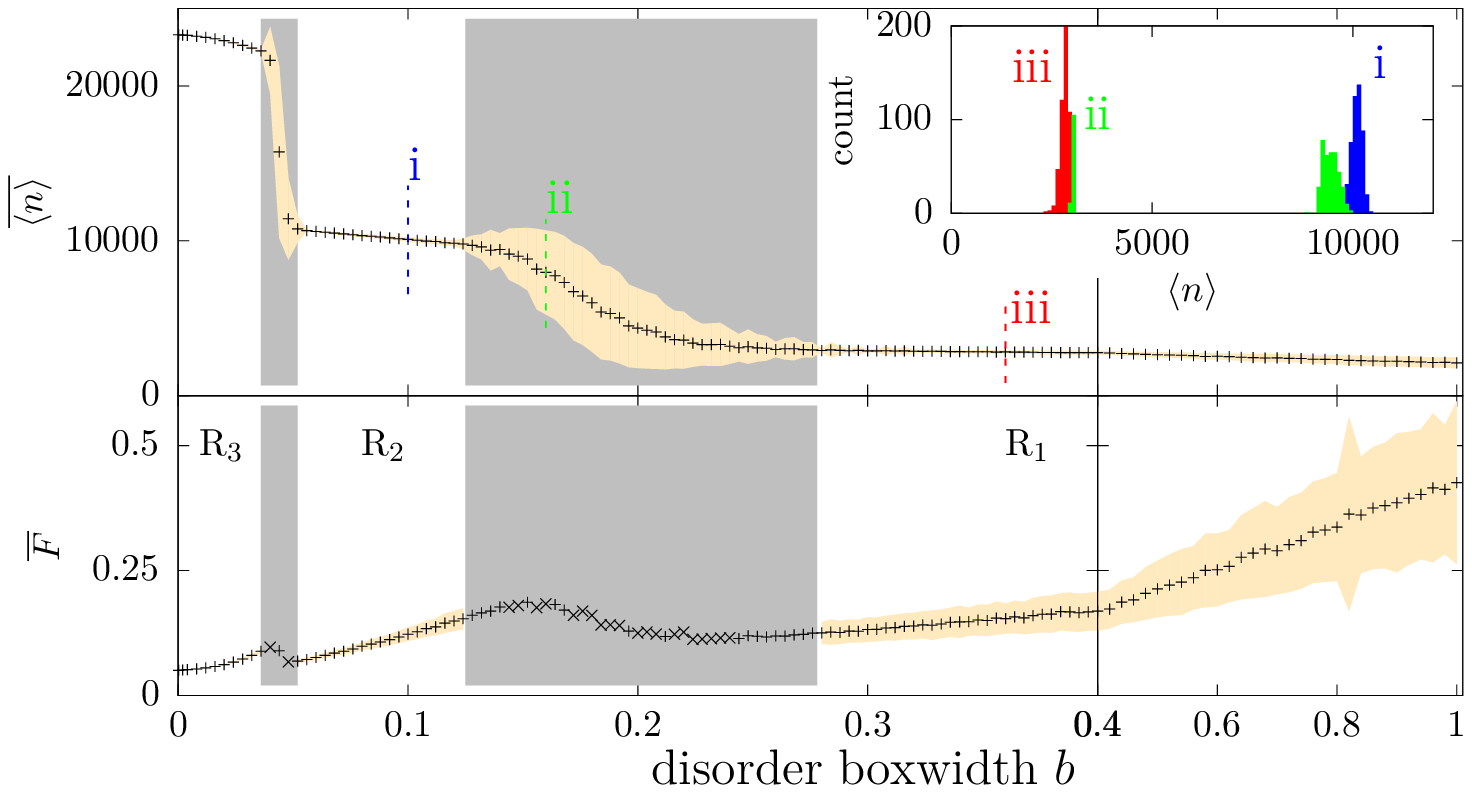}
		\label{fig:DisorderGateCharge:GSmall}
	}
	\subfigure[~$\mathcal{G}=0.07$]{
		\includegraphics[width=.48\textwidth]{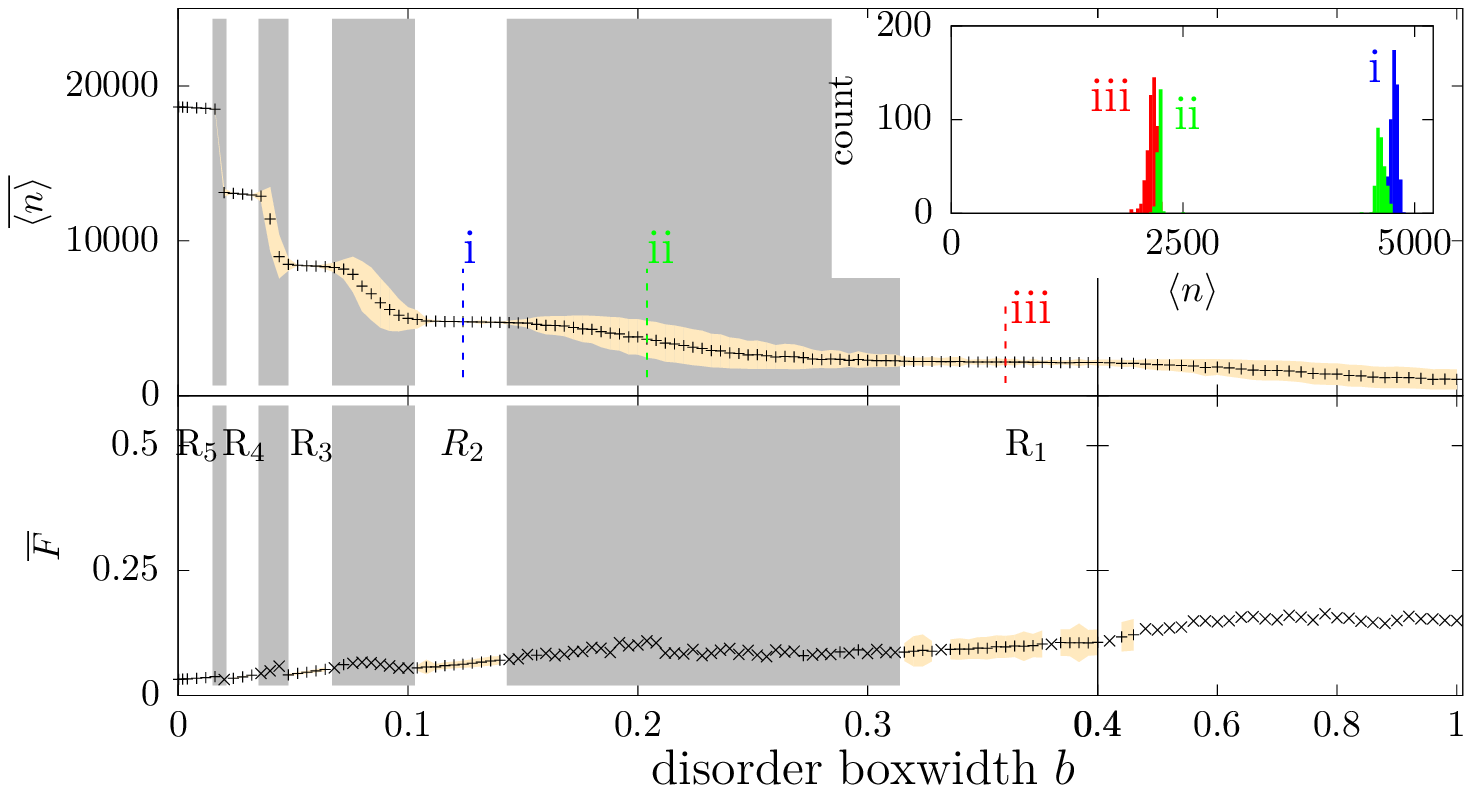}
		\label{fig:DisorderGateCharge:GLarge}
	}
	\caption{
		Mean photon-number expectation value $\overline{\erw{n}}$ and mean Fano factor $\overline{F}$ of a $M = 200$ atom lasing setup as a function of the boxwidth $b$ of a box distribution of the disordered atomic gate charges $N_{\mathrm{G},j}$.
		We average over $N_\mathrm{runs} = 500$ randomly generated systems.
		In the limit $b \to 0$ the systems correspond to the parameters marked by the white points in Fig.~\ref{fig:Squeezing}.
		Insets: Histogram of the distribution of the photon-number expectation value $\erw{n}$ at three different disorder 
		boxwidths $b$, marked by the dashed lines in the main plots.
		Plot parameters are $\kappa = 5 \times 10^{-5}\,\abs{\omega_\mathrm{eff}}$, $\overline{N}_\mathrm{G} = 0.5$, $E_{\mathrm{JL}} = \hbar \abs{\omega_\mathrm{eff}}$, $E_\mathrm{C} = 3.3 \,\hbar \abs{\omega_\mathrm{eff}}$, 
		$E_\mathrm{JU} = 0.5\,\hbar\abs{\omega_\mathrm{eff}}$, $\Gamma_\downarrow = 0.06\,\abs{\omega_\mathrm{eff}}$, and $\Gamma_\varphi^* = 0.013\,\abs{\omega_\mathrm{eff}}$. 
		The bare resonator frequency is $\omega_0 = 1.32\,\abs{\omega_\mathrm{eff}}$. 
		The regimes $\mathrm{R}_1$ to $\mathrm{R}_5$ and the different plot markers are discussed in the main text.
	}	
	\label{fig:DisorderGateCharge}
\end{figure}

Numerical results for a system consisting of $M=200$ artificial atoms and two different values of the parameter $\mathcal{G}$ are shown in Fig.~\ref{fig:DisorderGateCharge}.
The faint yellow areas indicate the standard deviation of $\erw{n}$ and $F$, respectively, which are a measure of the sample-to-sample variations of these quantities around their mean values $\overline{\erw{n}}$ and $\overline{F}$.
The limit $b \to 1$ corresponds to a spread of the gate charges over the entire possible range of values, i.e., maximum disorder.
The limit of a clean system, $b \to 0$, corresponds to the parameters of the ordered lasing system indicated by the white points in Fig.~\ref{fig:Squeezing}, i.e., the photon number is trapped at the third and the fifth root of the coupling matrix elements, respectively.

For this set of parameters we observe a non-monotonous behavior of the Fano factor as a function of the disorder boxwidth.
We discriminate different regimes,  in which the mean photon-number expectation value $\overline{\erw{n}}$ and the mean Fano factor $\overline{F}$ depend differently on the disorder boxwidth $b$. 
They are marked in Fig.~\ref{fig:DisorderGateCharge} by gray background boxes and labels $\mathrm{R}_1$ to $\mathrm{R}_5$.
To understand this behavior it is important to notice that an increase of the disorder boxwidth effectively decreases the number of atoms in the lasing setup. 
In the case of conventional lasing in the presence of disorder, this phenomenon has been discussed in detail in Ref.~\onlinecite{KMS_Disorder}. 
Atoms with a too large detuning or a too small coupling strength cannot participate in the lasing process any more and, therefore, the effective pumping strength $M \Gamma_\downarrow$ decreases. 
If the conventional photon number $n_\mathrm{c}$ corresponding to this effective pumping strength becomes lower than a certain root of the coupling matrix elements, the laser switches from a state trapped at this root to a state trapped at the next-lower root. 
Therefore, we observe plateaus in the mean photon-number expectation value $\erw{n}$ that correspond to the positions of the roots of the coupling matrix elements (label $\mathrm{R}_j$ for the $j$-th root). 

Between these trapped regions there are ranges of the disorder boxwidth where the photon number is either trapped at the higher or the lower root of the coupling matrix element, depending on the actual distribution of the gate charges (marked by a gray background box). 
The histograms of the photon-number expectation value $\erw{n}$ in the insets of Fig.~\ref{fig:DisorderGateCharge} show two distinct peaks corresponding to trapping at the two adjacent roots (data sets ``ii'').
In these transition regions, the mean photon-number expectation value $\overline{\erw{n}}$ and its standard deviation reflect a transition from a single-peaked distribution at a large photon number (data sets ``i'') via a double-peaked distribution to a single-peaked distribution at a small photon number (data sets ``iii''), rather than characterizing the mean photon number and its sample-to-sample fluctuations to be expected for a particular realization of the lasing setup.

Likewise, in the transition regions the mean Fano factor and its standard deviation are no longer well-defined quantities: 
If the lasing parameters happen to describe a lasing state that has a double-peaked photon statistics at both roots, a very large Fano factor $F \ggg 1$ is obtained. 
The data set ``ii'' in the inset of Fig.~\ref{fig:DisorderGateCharge:GSmall} shows one such event at $\erw{n} \approx 8800$. 
Therefore, the histogram of the Fano factor has not only two peaks at $F \lesssim 1$, corresponding to systems trapped at the two adjacent roots of the coupling, but also a tail of very rare realizations of lasing systems that have a Fano factor $F \ggg 1$. 
Due to these very rare events, $\overline{F}$ and its standard deviation are no longer smooth functions of the disorder 
boxwidth and we do not plot the Fano factor standard deviation in the transition regions as it lacks clear physical meaning. 
If we calculate the mean Fano factor $\overline{F}$ considering only those realizations of lasing systems with a Fano factor smaller than $F_\mathrm{cutoff} = 5$ a smooth dependence of this modified mean Fano factor of the disorder boxwidth is obtained. 
X-shaped data points indicate that lasing systems with very large Fano factors $F > F_\mathrm{cutoff}$ have been neglected.

Within the trapped regimes $\mathrm{R}_2$ and $\mathrm{R}_1$ in Fig.~\ref{fig:DisorderGateCharge:GSmall} and within all regimes $\mathrm{R}_5$ to $\mathrm{R}_1$ in Fig.~\ref{fig:DisorderGateCharge:GLarge} the mean photon-number expectation value decreases linearly as a function of the disorder boxwidth $b$, but the slope is quite small (note the change of the $b$ axis scale at $b = 0.4$). 
Likewise, the mean Fano factor $\overline{F}$ increases linearly with the disorder.
In the regime $\mathrm{R}_3$ in Fig.~\ref{fig:DisorderGateCharge:GSmall} a crossover from constant photon number (Fano factor) in the limit of very small disorder to a linear decrease (increase) for large disorder is observed.
In the trapped regimes $\mathrm{R}_j$ the sample-to-sample fluctuations of the photon-number expectation value and the Fano factor can be decreased by using a larger number of atoms $M$ and a larger resonator decay rate $\kappa$, while keeping the cooperativity $C$ and $n_\mathrm{c}$ constant. 
However, this is only a weak effect and there are constraints on the maximum possible number of atoms in the setup. 
For instance, the fluctuations at $b = 0.11$  are reduced by a factor of $0.6$ if a five times larger system is used. 

The weak dependence of $\overline{\erw{n}}$ and $\overline{F}$ on the disorder in the regions $\mathrm{R}_j$ makes them promising for an experimental realization of the CPT lasing setup. 
For the parameters considered in Fig.~\ref{fig:DisorderGateCharge:GSmall} the setup is expected to create photon-number squeezed resonator states characterized by a Fano factor $F \lesssim 0.5$ at a photon number of $\erw{n} \gtrsim 2100$ even in the presence of maximum charge disorder. 

Generally, the photon-number expectation value in the limit of maximum disorder, $b \to 1$, is determined by the position of the first root of the coupling matrix elements. 
Therefore, choosing a smaller value of $\mathcal{G}$ increases the brightness of the laser because the position of the first root is shifted towards larger photon numbers.
On the other hand, this increases the Fano factor in the limit of maximum disorder because the regime of conventional lasing for photon numbers below the first root of the coupling (cf.\ Fig.~\ref{fig:Squeezing}) increases and some atoms will eventually reach this regime of conventional lasing.
Therefore, there is a tradeoff between a high photon number and a small Fano factor in the limit $b \to 1$. 

Lasing at higher roots of the coupling matrix elements, e.g., for rather large values $\mathcal{G} \to 0.1$, yields
a photon-number squeezed statistics as well. 
However, setups with a large value of $\mathcal{G}$ are expected to be less robust against disorder,
because the number of transition regions between regimes trapped at different roots of the coupling increases and the ranges of the disorder boxwidth $b$ decrease, where a trapped state is robust against disorder (compare both subfigures of Fig.~\ref{fig:DisorderGateCharge}).

\section{Conclusions}\label{sec:Conclusions}
In this work we presented a lasing setup based on multiple Cooper-pair transistors coupled to a single microwave resonator. 
Over a broad range of experimentally feasible parameters this laser creates photon-number squeezed light at large intensity.
The photon-number squeezing arises because of the presence of roots of the coupling matrix elements between the charge states of the Cooper-pair transistor and the resonator. 

Compared to the systems discussed in Ref.~\onlinecite{KM_Squeezing}, based on two-level systems with longitudinal coupling
to a resonator, the Cooper-pair injection lasing setup considered here has two advantages:
First, it is very robust against the charge noise of the Cooper-pair transistor. 
We found a regime where the Fano factor is significantly smaller than unity and almost independent of the disorder strength, even in the presence of maximum charge noise and strong qubit dephasing.
Therefore, an experimental realization of this effect is very feasible.
Second, the realizable Fano factor is not bound from below by the number $M$ of artificial atoms, i.e., the emission intensity can be increased independently of the Fano factor by adding more Cooper-pair transistors to the setup.

Our results help the design and construction of bright and robust miniaturized sources of nonclassical microwave radiation, which
may have important applications in low-temperature experiments.
In particular, photon-number squeezed light can be useful for, e.g.~frequency-modulation spectroscopy
and, due to its well defined power, calibration of low-temperature devices.

\section*{Acknowledgements}
We acknowledge fruitful discussions with G.\ Sch\"on. 
This work has been funded from the DFG Grant No. MA 6334/3-1 and we acknowledge financial support by the Swiss SNF and the NCCR Quantum Science and Technology. 

\appendix
\section{Effect of direct couplings}
\label{app:Justifications}
In this Appendix we show that the ic coupling~\eqref{eqn:AdditionalCouplings:IC} between different CPTs, which so far had been neglected in the effective lasing Hamiltonian~\eqref{eqn:Heff}, is only a small perturbation of the lasing state if the total dephasing rate is larger than the coupling energies, $E_{\mathrm{ic},j,l} \ll \hbar\Gamma_{\varphi,j}$. 

Above the lasing threshold the expectation values $\erw{\sigma_z^j}$ and $\erw{\sigma_\pm^j} = s_\pm^j e^{\pm i \abs{\omega_\mathrm{eff}} t}$ can be calculated using the so-called semiclassical lasing theory \cite{SemiclassicalLaserTheoryMandel,SemiclassicalLaserTheoryWeidlich,SemiclassicalLaserTheoryAndre}. 
Within a rotating wave approximation it yields the following equations of motion for the classical amplitudes $s_z^j = \erw{\sigma_z^j}$ and $s_\pm^j$,
\begin{align}
	\frac{\d}{\d t} s_-^j = 
		&- (\Gamma_{\varphi,j} + i \Delta_j) s_-^j - i g_j s_z^j A^* \\
		&- 4 i \sum_{\stackrel{l \neq j}{l=1}}^M G_{z;l,j} s_z^l s_-^j + 2 i \sum_{\stackrel{l \neq j}{l=1}}^M G_{x;l,j} s_-^l s_z^j \comma \nonumber \\
	\frac{\d}{\d t} s_z^j = 
		&- \Gamma_{1,j} s_z^j - 2 i g_j \left( s_-^j A - s_+^j A^* \right) \\
		&+ 4 i \sum_{\stackrel{l \neq j}{l=1}}^M G_{x;l,j} \left( s_+^l s_-^j - s_-^l s_+^j \right) + \Gamma_{\uparrow,j} - \Gamma_{\downarrow,j} \comma \nonumber
\end{align}
where we used the abbreviations $\Delta_j = \epsilon_j/\hbar - \abs{\omega_\mathrm{eff}}$, $A = e^{i \omega_\mathrm{eff} t} \erw{e^{i \mathcal{G} (a + a^\dagger)}}$, $g_j = \sin^2 (\theta_j/2) E_{\mathrm{JU},j}/(2 \hbar)$, $G_{x;l,j} = \sin (\theta_j) \sin (\theta_l) E_{\mathrm{ic},j,l}/(2 \hbar)$, and $G_{z;l,j} = \cos (\theta_j) \cos (\theta_l) E_{\mathrm{ic},j,l}/(2 \hbar)$. 
We now expand the stationary solutions of $s_z^j$ and $s_\pm^j$ in a Taylor series in terms of $G_{x/z;l,j}$ and obtain the following first-oder corrections:
\begin{align*}
	[s_z^j]^{(1)} &\propto [s_z^j]^{(0)} \sum_{\stackrel{l \neq j}{l=1}}^M [s_z^l]^{(0)} \left( 8 \frac{\Delta_j G_{z;l,j}}{\Gamma_{\varphi,j}^2 + \Delta_j^2} - 4 \frac{g_l}{g_j} \frac{\Delta_l G_{x;l,j}}{\Gamma_{\varphi,l}^2 + \Delta_l^2} \right) \comma \\
	[s_-^j]^{(1)} &\propto [s_z^j]^{(0)} \sum_{\stackrel{l \neq j}{l=1}}^M [s_z^l]^{(0)} \Big( \begin{aligned}[t] 
		&2 \frac{G_{x;l,j}}{\Gamma_{\varphi,j} + i \Delta_j} \frac{g_l}{\Gamma_{\varphi,l} + i \Delta_l} \\
		- &4 \frac{G_{z;l,j}}{\Gamma_{\varphi,j} + i \Delta_j} \frac{g_j}{\Gamma_{\varphi,j} + i \Delta_j} \Bigg) 
	\end{aligned} \\
	&\quad + \mathcal{O}\left( [s_z^j]^{(1)} \right) \comma
\end{align*}
where $[s_{z/-}^j]^{(n)}$ denotes the $n$-th order coefficient of the Taylor expansion of $s_{z/-}^j$.
The omitted prefactors are of the order of unity. 
Typical parameters of the CPT lasing setup are $E_{\mathrm{C},j} \gtrsim E_{\mathrm{JU},j} \approx \hbar \Gamma_{\varphi,j}$, $\Delta \approx 0$, $\tilde{C} \gg C_{\Sigma,j}$, and, therefore, $E_{\mathrm{C},j} \gg E_{\mathrm{ic},l,j}$. 
Hence, the corrections $[s_z]^{(1)}$ and $[s_-^j]^{(1)}$ due to CPT coupling (ic) terms are suppressed by the factors $\Delta_j/\Gamma_{\varphi,j} \ll 1$ and 
\begin{align*}
	\frac{G_{x/z;l,j}}{\Gamma_{\varphi,j}} \propto \frac{E_{\mathrm{ic},j,l}}{E_{\mathrm{C},j}} \ll 1 
\end{align*}
compared to the results obtained by neglecting the CPT coupling (ic) terms. 
Intuitively, weak interatomic couplings in a multiatom lasing setup level differences in the population inversion of the individual atoms, which may arise because of fluctuations in the pumping or coupling strength, but they do not spoil lasing processes. 

\section{Total lasing Hamiltonian in the rotating wave approximation}
\label{sec:FullHamiltonian}
In this Appendix we give the full form of the effective lasing Hamiltonian~\eqref{eqn:Heff}, taking into account the terms originating from the CPT coupling (ic) term. 
Having restricted the full circuit Hamiltonian~\eqref{eqn:Hcircuit} including the interaction term~\eqref{eqn:AdditionalCouplings:IC} to the two lowest energy eigenstates of each CPT and having applied a rotating-wave approximation, we obtain
\begin{align*}
	\tilde{H}_\mathrm{eff} &= \hbar \omega_\mathrm{eff} a^\dagger a + \sum_{j=1}^M \frac{\tilde{\epsilon}_j}{2} \sigma_z^j \\
	&+ \sum_{j=1}^M \sum_{n=0}^\infty \left( \hbar A_{n+1,n}^j \sigma_+^j \ket{n+1} \bra{n} + \mathrm{h.c.} \right) \\
	&+ \sum_{j=1}^M \sum_{\stackrel{l \neq j}{l=1}}^M \frac{E_{\mathrm{ic},j,l}}{2} \left( \cos \tilde{\theta}_j \cos \tilde{\theta}_l \sigma_z^j \sigma_z^l + \sin \tilde{\theta}_j \sin \tilde{\theta}_l \sigma_x^j \sigma_x^l \right) \comma
\end{align*}
where the effective resonator frequency is given by Eq.~\eqref{eqn:EffectiveFrequency} and the level-splitting energy and mixing angles are 
\begin{align}
	\tilde{\epsilon}_j &= \sqrt{\left[ 4 E_{\mathrm{C},j} (1 - 2 N_{\mathrm{G},j}) + \delta \epsilon_j \right]^2 + E_{\mathrm{JL},j}^2} \comma
	\label{eqn:epsilonTotal} \\
	\tan \tilde{\theta}_j &= - \frac{E_{\mathrm{JL},j}}{4 E_{\mathrm{C},j} (1 - 2 N_{\mathrm{G},j}) + \delta \epsilon_j} \comma \\
	\delta \epsilon_j &= 2 \sum_{\stackrel{l \neq j}{l = 1}}^M E_{\mathrm{ic},j,l} \fullstop
\end{align}
Note that $\delta \epsilon_j$ shifts the sweet spot of the charge qubit, but it is always possible to tune the gate voltages $V$ and $W_j$ accordingly to compensate for this effect.

\end{document}